\documentclass[12pt,twoside]{article}
\usepackage{amssymb}
\usepackage{amsmath}
\usepackage{mathabx}
\usepackage{latexsym}
\usepackage{longtable}
\usepackage{epsfig}
\usepackage{graphicx,bbm,psfrag}
\graphicspath{{images/}}
\usepackage{cite}

\DeclareMathAlphabet{\mathsfit}{T1}{\sfdefault}{\mddefault}{\sldefault}
\SetMathAlphabet{\mathsfit}{bold}{T1}{\sfdefault}{\bfdefault}{\sldefault}

\begin{document}
\begin{titlepage}
\vspace*{3cm}
\begin{center}
\Large{\textbf{Hydrodynamic Equations  for the Ablowitz-Ladik \smallskip\\ 
Discretization
of the Nonlinear Schr\"{o}dinger Equation}}\bigskip\bigskip\bigskip
\end{center}
\begin{center} 
{\large{Herbert Spohn}}\bigskip\bigskip\\
Departments of Mathematics and Physics, Technical University Munich,\smallskip\\
Boltzmannstr. 3, 85747 Garching, Germany
\end{center}
\vspace{3cm}
\textit{Dedication}: Freeman Dyson pioneered the investigation of random matrices. For a glimpse on the early founding period of the subject one can consult his introductory chapter in the 2011 Oxford Handbook on Random Matrices. Our  
contribution is related, since we elucidate a close connection between the high temperature limit of beta CUE and the Ablowitz-Ladik discretized version of the nonlinear 
Schr\"{o}dinger equation in one spatial dimension. In deep appraisal of an outstanding scientist, this article is dedicated to Freeman Dyson. 
\begin{flushright}
October 20, 2021
\end{flushright}
\vspace{2.8cm} 
\textbf{Abstract}:  Ablowitz and Ladik discovered a discretization which preserves the 
integrability of the nonlinear Schr\"{o}dinger equation in one dimension. 
We compute the generalized free energy of this model and determine the GGE averaged fields and their currents.
They are linked to the mean-field circular unitary matrix ensemble (CUE).
The resulting hydrodynamic equations follow the pattern already known from other integrable many-body systems.
Studied is also the discretized modified Korteweg-de-Vries equation which turns out to be related to the beta Jacobi 
log gas.


\end{titlepage}
\section{Introduction}
\label{sec1}

A famous integrable classical field theory in $1+1$ dimensions is the nonlinear Schr\"{o}dinger equation. 
In the defocusing case the wave field, $\psi(x,t) \in \mathbb{C}$, is governed by 
\begin{equation} 
\label{1.1}
\mathrm{i}\partial_t\psi = - \partial_x^2\psi + 2|\psi|^2 \psi.
\end{equation}
While many properties of this equation have been studied \cite{ABT04}, given the more recent interest in generalized hydrodynamics
\cite{CDY16,BCDF16,D19a,JSTAT21}
our goal is to investigate the Euler type spacetime scale for this nonlinear wave equation.

In hydrodynamics, one considers random initial data with an energy far above the ground state energy. The resulting physical picture
is based on the notion of local equilibrium. In a small cell, still containing many particles, the system is  in one of its 
equilibrium states. In approximation, the equilibrium parameters  are changing slowly on the scale set by the inter-particle distance and evolve according to a system of coupled hyperbolic conservation laws.
In a conventional strongly interacting fluid in one dimension, equilibrium is labelled by three parameters. In strong contrast, for integrable systems the time-stationary states require an extensive number of parameters.   Thus the first step in any  derivation of hydrodynamic equations consists of a detailed study of  stationary states.

For NLS the  densities of the locally conserved fields are known, see \cite{GK14} and references therein. The beginning of the list reads
\begin{equation}
\label{1.2}
Q^{[0]}(x) = |\psi(x)|^2,
\qquad Q^{[1]}(x) = -\mathrm{i} \bar{\psi}(x)\partial_x \psi(x),
\qquad Q^{[2]}(x) = |\partial_x \psi(x)|^2 +  |\psi(x)|^4.
\end{equation}
For a given bounded interval $\Lambda \subset \mathbb{R}$, the total conserved quantities then become
\begin{equation}
\label{1.3}
Q_\Lambda^{[n]} =
\int_\Lambda \hspace{-3pt}\mathrm{d}x Q^{[n]}(x), \qquad n = 0,1,....\, .
\end{equation}
Formally, the time-stationary generalized Gibbs ensembles (GGE) are of the form
\begin{equation}
\label{1.4}
\exp\Big[ - \sum_{n=0}^\infty \mu_n Q_\Lambda^{[n]}\Big] \prod_{x \in \Lambda}\mathrm{d}^2 \psi(x),
\end{equation}
where the $\mu_n$'s are suitable chemical potentials. For the thermal case, $n=0,1,2,$ much work has been invested to construct  a proper probability measure, see \cite{LRS88,B94,OQ13,FKSS17} for a very partial account. A class of generalized Gibbs measures has been  constructed
in \cite{Z01}. While the precise statement  in \cite{Z01} is more complicated, the basic idea is to restrict the sum to some highest even $n$
and to use as reference measure the Gaussian measure with an energy
\begin{equation}
\label{1.5}
Q_\Lambda^{[0]} +    \int_\Lambda \hspace{-3pt}\mathrm{d}x |\partial_x^{n/2} \psi(x)|^2  . 
\end{equation} 
The technical part is to establish that, for an appropriate choice of chemical potentials, the exponential of all lower order terms can be integrated with respect to the Gaussian measure. Very roughly the such constructed measure is concentrated on $(n-2)$ times differentiable functions. Time-stationarity
is established. To go beyond such an existence result seems to be a difficult problem.

In numerical simulations of NLS one  discretizes the equation. While generically this would break integrability, surprisingly enough, in many cases there is one very specific discretization for which  integrability is maintained. For NLS such a discretization was discovered 
by Ablowitz and Ladik  \cite{AKNS74,AL75,AL76}. For convenience we will use here AL as acronym (rather than IDNLS as in \cite{ABT04}) and our contribution is focussed on the Ablowitz-Ladik system. 

A further example in the same spirit is the classical sinh-Gordon equation, 
\begin{equation}
\label{1.6}
\partial_t^2\phi - \partial_x^2\phi + \sinh \phi = 0
\end{equation}
with $\phi(x,t)$ a real-valued wave-field. In \cite{BDWY18} the hydrodynamic equations of this nonlinear wave equation are derived and studied. The  reported numerical simulations are based on an analytic continuation of the discretized sine-Gordon equation discovered in \cite{O78}. 
A further case is the integrable Landau-Lifshitz model of a one-dimensional magnet. Here the spin field is a three-vector, $\vec{S}(x,t)$, with 
$|\vec{S}| = 1$ and the equations of motion are
\begin{equation}
\label{1.7}
\partial_t\vec{S} = \vec{S} \times \partial_x^2 \vec{S} + \vec{S}\times \mathsf{J} \vec{S}, \quad \mathsf{J} = \mathrm{diag}(0,0,\delta).
\end{equation}
The naive discretization is non-integrable. For recent investigations we refer to \cite{GMI19,DDD20} and references therein. The integrable discretization was discovered in \cite{T77,S79}, see also \cite{FT87}.  A numerical simulation of the integrable chain is reported in \cite{DKSA19}.

As we will see, the AL lattice equations are structurally rather similar to the classical Toda lattice, for which fairly detailed notes 
are available \cite{S21}.  When pointing out such similarity we merely refer to these notes. Still our text is essentially self-contained and the reader may as well just ignore the cross links.

As other integrable wave equations, the NLS admits soliton solutions. Rather than assuming an initial state which is locally GGE, 
random initial conditions can be imposed in the form of a solitons gas,
meaning that location and velocity of a soliton are random, similar to a classical gas of point particles. In approximation, on large scales the time evolution of 
such a gas can described by a coupled set of hyperbolic conservation laws. For more details recommended is a very recent review 
article by G. El \cite{E21}, who uses the Korteweg-De Vries equation as his most prominent example.

\section{The periodic AL system}
\setcounter{equation}{0}
\label{sec2}
Upon discretization, the wave field is over the one-dimensional lattice $\mathbb{Z}$, $\psi_j(t) \in \mathbb{C}$,
 and is governed by   
\begin{equation}\label{2.1}
\mathrm{i}\frac{d}{dt}\psi_j  = - \psi_{j-1}  + 2 \psi_j - \psi_{j+1} +  |\psi_j|^2 (\psi_{j-1} + \psi_{j+1}).
\end{equation}
 Hence
\begin{equation} \label{2.2} 
\mathrm{i}\frac{d}{dt}\psi_j  = - (1 - |\psi_j|^2)(\psi_{j-1} + \psi_{j+1}) + 2 \psi_j,
\end{equation}
Setting $\alpha_j(t) = \mathrm{e}^{2\mathrm{i}t}\psi_j(t)$, one arrives at the standard version
\begin{equation}\label{2.3} 
\frac{d}{dt}\alpha_j  = \mathrm{i} \rho_j^2 (\alpha_{j-1} + \alpha_{j+1}),\quad \rho_j^2 = 1 - |\alpha_j|^2.
\end{equation}
Clearly the natural phase space is $\alpha_j \in \mathbb{D}$ with the unit disk $\mathbb{D} = \{z||z| \leq 1\}$. In principle, whenever $\alpha_j (t)$
hits the boundary of $\mathbb{D}$, it freezes and thereby decouples the system. As we will discuss, a conservation law ensures that, if initially away from the boundary, the solution will stay so forever.

Our main focus is the generalized free energy, for which the standard set-up is a finite ring of $N$ sites, labelled as $j = 0,...,N-1$ with periodic boundary conditions,
$\alpha_{j+N} = \alpha_j$. While the periodic system is integrable, there seems to be no method for obtaining its generalized free energy 
 in the limit $N \to \infty$. However,  employing a judicious choice of boundary conditions 
such a task becomes feasible. Therefore we have to discuss separately  the ring and a segment with boundary conditions.\bigskip\\
\textit{Conserved fields}. We consider a ring of $N$ sites.  The evolution equation are of hamiltonian form by 
regarding $\alpha$ and its complex conjugate $\bar{\alpha}$ as canonically conjugate variables and introducing
the weighted Poisson bracket
\begin{equation}\label{2.4} 
\{f,g\}_\mathrm{AL} =\mathrm{i} \sum_{j=0}^{N-1} \rho_j^2\Big(\frac{\partial f}{\partial{\bar{\alpha}_j}} 
\frac{\partial g}{\partial{\alpha_j}}   - \frac{\partial f}{\partial{\alpha_j}} 
\frac{\partial g}{\partial{\bar{\alpha}_j}} 
\Big).
\end{equation}
The Hamiltonian of the AL system reads
\begin{equation}\label{2.5} 
H_N = - \sum_{j=0}^{N-1} \big( \alpha_{j-1} \bar{\alpha}_{j} + \bar{\alpha}_{j-1} \alpha_{j}\big).
\end{equation}
One readily checks that indeed
\begin{equation} \label{2.6}
\frac{d}{dt}\alpha_j = \{\alpha_j,H_N\}_\mathrm{AL}   = \mathrm{i} \rho_j^2 (\alpha_{j-1} + \alpha_{j+1}) .
\end{equation}

The next step is to find out the locally conserved fields.  Guided by other integrable models, the convenient tool is a Lax matrix, if available. I. Nenciu \cite{N05,N06}
discovered that this role is played by a  
Cantero-Moral-Vel\'{a}zquez (CMV) matrix \cite{CMV03,CMV05,KN07,S07}. The basic building blocks are the $2\times 2$ matrices,
which  requires $N$ to be \textit{even} because of periodic boundary conditions. One defines
\begin{equation}\label{2.7} 
\Xi_j =
\begin{pmatrix}
\bar{\alpha}_j & \rho_j \\
\rho_j & -\alpha_j  \\
\end{pmatrix}
\end{equation}
and forms the $N\times N$ matrices
\begin{equation}\label{2.8} 
 L_N = \mathrm{diag}(\Xi_0,\Xi_2,\dots, \Xi_{N-2}), \quad \big( M_N\big)_{i,j = 1,...,N-2} = \mathrm{diag}(\Xi_1,\Xi_3,\dots, \Xi_{N-3}),
\end{equation}
together with $(M_N)_{0,0} =  -\alpha_{N-1}$, $(M_N)_{0,N-1} = \rho_{N-1}$, $(M_N)_{N-1,0} =  \rho_{N-1}$, $(M_N)_{N-1,N-1} = \bar{\alpha}_{N-1}$. More pictorially $L_N$ corresponds to the blocking $(0,1),...,(N-2,N-1)$, while $M_N$ uses the by $1$ shifted blocking $(1,2),...,(N-1,0)$. The CMV matrix associated to the coefficients $\alpha_0,\dots,\alpha_{N-1}$ is then given by 
\begin{equation} \label{2.9}
C_N = L_N M_N.
\end{equation}
Obviously, $L_N,M_N$ are unitary and so is $C_N$. The  eigenvalues of $C_N$ are denoted by $\mathrm{e}^{\mathrm{i}\vartheta_j}$, $\vartheta_j \in [0,2 \pi]$,
$j = 1,\dots,N$.  Of course, the eigenvalues depend on $N$, which is suppressed in our notation.

Next we define for a general matrix, $A$, the $+$ operation as
\begin{equation} \label{2.10}
 (A_+)_{i,j} = \begin{cases} A_{i,j}  &\mathrm{if} \,\,i < j,\\
   \tfrac{1}{2}A_{i,j} &\mathrm{if} \,\,i = j,\\
  0 &\mathrm{if} \,\,i > j.
 \end{cases}
\end{equation}
Then one version of the Lax pair reads
\begin{equation}\label{2.11}
\{C_N,\mathrm{tr}(C_N)\}_\mathrm{AL} = \mathrm{i} [C_N,(C_N)_+],\qquad \{C_N,\mathrm{tr}(C_N^*)\}_\mathrm{AL} = \mathrm{i} [C_N,(C_{N+})^*].
\end{equation}
Since the Poisson bracket acts as a derivative, one deduces
\begin{equation} \label{2.12}
\{(C_N)^n,\mathrm{tr}(C_N)\}_\mathrm{AL} = \sum_{m=0}^{n-1} (C_N)^m\mathrm{i}[C_N,C_{N+}](C_N)^{n-m-1} = \mathrm{i} [(C_N)^n,C_{N+}], 
\end{equation}
and similarly 
\begin{equation} \label{2.13}
\{(C_N)^n,\mathrm{tr}(C_N^*)\}_\mathrm{AL} =  \mathrm{i} [(C_N)^n,(C_{N+})^*]. 
\end{equation}
Hence the locally conserved fields are given by
\begin{equation} \label{2.14}
Q^{[n],N} = \mathrm{tr}\big[(C_N)^n\big].
\end{equation}
By a similar argument, it can be shown that the mutual Poisson brackets vanish,
\begin{equation} \label{2.15}
\{Q^{[n],N}, Q^{[n'],N}\}_\mathrm{AL} = 0 \bigskip.
\end{equation}

The fields $Q^{[n],N}$ are complex-valued. Physically real-valued phase function are preferred, which is achieved through taking
real and imaginary parts,
\begin{eqnarray}\label{2.16} 
&& Q^{[n,+],N} = \tfrac{1}{2}\mathrm{tr}\big[(C_N)^n + (C_N^*)^n\big] = \mathrm{tr}\big[\cos((C_N)^n)\big]\nonumber\\
&& Q^{[n,-],N} = -\tfrac{1}{2}\mathrm{i}\,\mathrm{tr}\big[(C_N)^n - (C_N^*)^n\big] = \mathrm{tr}\big[\sin((C_N)^n)\big],
\end{eqnarray}
with $n = 1,\dots, N/2$. These fields have a density, respectively given by
\begin{equation} \label{2.17}
Q_j^{[n],N}= Q_j^{[n,+],N}+\mathrm{i}Q_j^{[n,-],N} = ((C_N)^n)_{j,j}.
\end{equation}
 Although the matrices $L_N,M_N$ have a basic $2\times 2$ structure,  the densities of the conserved fields are shift covariant by $1$.
 Let us  introduce the left shift, $\tau$, of the sequence $\alpha = (\alpha_0,\dots,\alpha_{N-1})$ by 
$(\tau \alpha)_j = \alpha_{j+1}$. Then, as established in Appendix A,
\begin{equation} \label{2.18}
Q^{[n,\sigma],N}_{j+1}(\alpha) = Q^{[n,\sigma],N}_j(\tau \alpha)
\end{equation}
with the convention $\sigma = \pm$.  

Later on, we will consider the infinite volume limit, $N \to \infty$. This will always be understood as a two-sided limit. For example the infinite volume limit of $L_N$,
denoted by $L$, is $L  = \mathrm{diag}(\dots, \Xi_{-2},\Xi_0,\Xi_2,\dots)$ and correspondingly $M  = \mathrm{diag}(\dots, \Xi_{-1},\Xi_1,\Xi_3,\dots)$.
$L,M$ are unitary operators on the Hilbert space $\ell_2(\mathbb{Z})$, and so is $C = LM$. The traces in \eqref{2.10} have no limit, but  densities do.
The matrix elements of $C^n$  can be expanded as the sum
\begin{equation}\label{2.19}  
(C^n)_{i,j} =  \sum_{j_1\in \mathbb{Z}}\dots \sum_{j_{2n-1}\in \mathbb{Z}} L_{i,j_1}M_{j_1,j_2} \dots L_{j_{2n-2},j_{2n-1}} M_{j_{2n-1},j},
\end{equation}
 which consists of a finite number of terms, only. For the infinite system the index $n$ runs over all positive integers.
 The infinite volume densities, $Q^{[n,\sigma]}_j$, are strictly local functions of $\alpha$ with support of at most $2n$ sites. The sum in \eqref{2.19} can be viewed
 as resulting from a nearest neighbor $2n$ step random walk from left to right. For this purpose one considers a checkerboard on $[0,2n] \times \mathbb{R}$. The unit square with corners $(0,0),(1,0),(1,1), (0,1)$ is white. Single steps of the walk are either horizontal, $j \leadsto j$, or up-down, $j \leadsto j\pm 1$. Such diagonal steps are permitted only on white squares.
 The matrix element $(C^n)_{i,j}$ is then the sum over all $2n$ step walks starting at $i$ and ending
 at $j$. Each walk represents a particular polynomial obtained by taking the product of local weights along the walk. The weights are\bigskip\\
 \begin{tabular}{ll}
 \hspace{30pt}$\rho_j$ &  for the diagonal steps $j \leadsto j+1$ and $j+1 \leadsto j$,\\
 \hspace{30pt}$\bar{\alpha}_{j}$ & for the horizontal step $j \leadsto j$  in case the lower square is black, \\
 \hspace{30pt}$-\alpha_{j-1}$ & for the horizontal step $j \leadsto j$  in case the upper square is black. \\
 \end{tabular} 
 \bigskip\\
  As examples, $C_{j,j} = - \alpha_{j-1}\bar{\alpha}_j $, $H_{j}  = C_{j,j} +\bar{C}_{j,j}$,  and $(C^2)_{j,j} =
 \alpha_{j-1}^2\bar{\alpha}_j^2 - \rho_{j-1}^2\alpha_{j-2}\bar{\alpha}_j  - \rho_{j}^2\alpha_{j-1}\bar{\alpha}_{j+1}$.
 Note that densities are not unique in general, while the total conserved fields, $Q^{[n],N}$, are unique. To illustrate, in the previous formula, an equivalent density would be $\alpha_{j-1}^2\bar{\alpha}_j^2 - 2\rho_{j}^2\alpha_{j-1}\bar{\alpha}_{j+1}$.
 
The CMV matrix misses one physically very important field, namely
\begin{equation}\label{2.20}
Q^{[0],N} =  - \sum_{j=0}^{N-1}  \log(\rho_j^2).
\end{equation}
To simplify notation we set $[0] = [0,\sigma]$ and $0 = 0\sigma$. The time-derivative of $Q^{[0],N}$ yields a telescoping sum, which vanishes on a ring. In lack of a common name we call $Q^{[0],N}$ the log intensity.
The log intensity vanishes for small amplitudes $|\alpha_j|^2$ and diverges at the maximal value, $|\alpha_j|^2 = 1$. Note also that
 \begin{equation}\label{2.21} 
\exp\big(-Q^{[0],N}\big) =  \prod_{j=0}^{N-1}\rho_j^2
\end{equation}
is conserved. Thus if initially $\exp\big(-Q^{[0],N}\big) > 0$, it stays so for all times, guaranteeing that the 
phase space boundary is never reached.\\\\
\textit{Generalized Gibbs ensemble}. Hydrodynamics is based on the propagation of local equilibrium. For the micro-canonical
equilibrium measure, the Statistical Mechanics rule is to adopt the uniform measure on the hypersurface defined through fixing the values of all conserved fields. For nonintegrable chains of $N$ sites, its codimension would be $1,2,3$, depending on the model. 
 As claimed by the integrable systems community, the Statistical Mechanics  rule applies even in case
of an extensive number conservation laws. In fact, in favorable  situations one can control the hamiltonian written in terms of action
variables. If this function has no flat pieces, then the uniform measure on invariant tori is approached in the long-time limit, almost surely. 
The AL system has a phase space of dimension $2N$.  The $Q^{[n,\sigma],N}$'s constitute $N$ conservation laws. Together with 
$Q^{[0],N}$, such a rule means the uniform measure  on an invariant torus of dimension $N-1$. 
As for other problems in statistical mechanics, more accessible is the grand-canonical version, a route also adopted here. One would expect that for large $N$
this hardly makes any difference, provided only averages of local observables are considered. In our context, to prove such an
equivalence of ensembles stays as one open problem.

The natural a priori measure of the AL model is the product measure
\begin{equation} \label{2.22}
\prod_{j=0}^{N-1}\mathrm{d}^2\alpha_j (\rho_j^2)^{P-1}= \prod_{j=0}^{N-1}\mathrm{d}^2\alpha_j (\rho_j^2)^{-1}\exp\big(-PQ^{[0],N}\big) 
\end{equation}
on $\mathbb{D}^N$. To normalize the measure, $P>0$ is  required.  The log intensity is controlled by the parameter $P$ which, in analogy to the 
Toda lattice, is called pressure. Small $P$ corresponds to maximal log intensity, i.e. $|\alpha_j|^2 \to 1$,  and large $P$ to low log intensity, i.e. $|\alpha_j|^2 \to 0$. 
In the grand-canonical ensemble, the Boltzmann weight is constructed from a linear combination of the conserved fields, which is written as
\begin{equation} \label{2.23}
 \sum_{n\in \mathbb{Z}}\mu_n \mathrm{tr}\big[(C_N)^n\big] = \mathrm{tr}\big[\hat{V}(C_N)\big], \qquad \hat{V}(z)=  \sum_{n\in \mathbb{Z}}\mu_n z^n.
\end{equation}
The chemical potentials, $\mu_n$, are assumed to be independent of $N$.  To have the trace real-valued one imposes $\mu_n = \bar{\mu}_{-n}$.
Also the normalization $\mu_0 = 0$  is adopted. Combining \eqref{2.22} and \eqref{2.23} then  yields the generalized Gibbs ensemble (GGE) as
\begin{equation} \label{2.24}
Z_{N}(P,V)^{-1}\prod_{j=0}^{N-1}\mathrm{d}^2\alpha_j (\rho_j^2)^{P-1}\exp\big(-\mathrm{tr}[\hat{V}(C_N)]\big), \qquad P >0 .
\end{equation}
$Z_{N}(P,V)$ is the normalizing partition function. As label, the more natural object turns out to be the Fourier transform of the sequence $\{\mu_n, n\in \mathbb{R}\}$,
\begin{equation} \label{2.25} 
V(w) = \sum_{n\in \mathbb{Z}}\mu_n \mathrm{e}^{\mathrm{i}nw} = \hat{V}(\mathrm{e}^{\mathrm{i}w}).
\end{equation}
$V$ is a real-valued function on $[0,2\pi]$. For the Toda lattice the corresponding object lives on $\mathbb{R}$ and is called confining potential 
because it confines the eigenvalues of the Lax matrix. For the CMV matrix the eigenvalues are on the unit circle and there is nothing to confine. 
To distinguish
from other potentials, for convenience we still stick to ``confining". 
If $\hat{V}$ is given by a finite sum, then the interaction of the Gibbs measure in \eqref{2.24} is of finite range. In this case the infinite volume limit can be
controlled through transfer matrix methods. In particular,  the limit measure is expected to have  a finite correlation length.
Presumably a larger set of confining potentials could be allowed, as studied in \cite{G21} for the Toda lattice. Finite volume expectations with respect to 
the measure in \eqref{2.24} are denoted by $\langle \cdot \rangle_{P,V,N}$ and their infinite volume limit by $\langle \cdot \rangle_{P,V}$.
The GGE is labelled by the pressure $P$ and  some smooth function on the unit circle.

The generalized free energy, $F_\mathrm{AL}$, is defined through 
\begin{equation} \label{2.26}
\lim_{N \to \infty} - \frac{1}{N} \log Z_{N}(P,V) = F_\mathrm{AL}(P,V).
\end{equation}
In the hydrodynamic context of particular interest is the empirical density of states (DOS),
\begin{equation} \label{2.27}
\rho_{Q,N}(w) \mathrm{d} w   = \frac{1}{N} \sum_{j=1}^N \delta (w - \vartheta_j) \mathrm{d}w    
\end{equation}
with the $\mathrm{e}^{\mathrm{i}\vartheta_j}$'s eigenvalues of $C_N$. $\rho_{Q,N}$ is a probability measure on $[0,2\pi]$ and has an almost sure limit as
\begin{equation} \label{2.28}
\lim_{N \to \infty} \rho_{Q,N}(w) = \rho_{Q}(w).
\end{equation}
To see the significance of the DOS, we first introduce  the trigonometric functions $\varsigma_0(w) =1$, $\varsigma_{n-}(w) = \sin(nw)$, and $\varsigma_{n+}(w) = \cos(nw)$, $n = 1,2,\dots$ . They span the Hilbert space $L^2([0,2\pi], \mathrm{d}w)$. Then, for the trigonometric moments of   $\rho_{Q,N}(w)$,
\begin{equation} \label{2.29}
 \langle \rho_{Q,N} \varsigma_{n\sigma} \rangle = N^{-1}\langle Q^{[n,\sigma],N}\rangle_{P,V,N}, \quad \lim_{N \to \infty} N^{-1}\langle Q^{[n,\sigma],N}\rangle_{P,V,N} = 
 \langle Q_0^{[n,\sigma]}\rangle_{P,V} = \langle \rho_{Q} \varsigma_{n\sigma} \rangle.
\end{equation}
Here $\langle \cdot \rangle$ is simply a short hand  for the integration over $[0,2\pi]$.
The limit value can also be expressed as variational derivative of the generalized free energy per site,
\begin{equation} \label{2.30}
\frac{ \mathrm{d}}{\mathrm{d}\kappa} F_\mathrm{AL}(P,V +\kappa \varsigma_{n\sigma})\big|_{\kappa = 0} = \langle Q_0^{[n,\sigma]}\rangle_{P,V}.
\end{equation}
In addition, one introduces for the average log intensity, denoted by $\nu$, for which
\begin{equation} \label{2.31}
\nu =  \langle Q_0^{[0]}\rangle_{P,V} = \partial_P F_\mathrm{AL}(P,V).
\end{equation}

For $V=0$, one readily obtains  $F_\mathrm{AL}(P,0)= \log(P/\pi)$ with log intensity $\nu(P) = P^{-1} > 0$. Hence there is no high pressure phase as known for the Toda lattice, see \cite{S21}, Section 8, and \cite{MS21}. Thermal equilibrium corresponds to  $V(w) = \beta \cos w$ with $\beta$ the inverse temperature.  Then the Gibbs measure in \eqref{2.24} has 
nearest neighbor interactions and explicit expressions seem no longer to be available, see however the note at the end of Section
\ref{sec5}.

While the existence of the infinite volume limit is reassuring,  more computable expressions are needed so to write down the hydrodynamic equations.
In demand would be the joint probability density for the eigenvalues and the resulting DOS. This looks difficult. Fortunately,  R. Killip and I. Nenciu  \cite{KN04}
discovered that through  a suitable modification of the boundary conditions the corresponding volume element can be transformed to only depend on the eigenvalues.
\section{Circular matrices with slowly varying pressure ramp}
\label{sec3}
\setcounter{equation}{0}
Following \cite{KN04} we modify the CMV matrix at the two boundaries. As before, the number  $N$ of sites is even. $L_N$ remains unchanged, $M_N$ is modified to  
$M_N^\diamond$, where $(M_N^\diamond)_{0,0} =1$, $(M_N^\diamond)_{0,N-1}= 0$, $(M_N^\diamond)_{N-1,0}=0$, and $(M_N^\diamond)_{N-1,N-1}
= \mathrm{e}^{\mathrm{i}\phi}$, $\phi \in [0,2\pi]$. This leads to the particular CMV matrix
\begin{equation} \label{3.1}
C_N^\diamond = L_NM_N^\diamond.
\end{equation}
For the a priori measure  \eqref{2.22} the pressure $P$ is constant, which is now modified to  a  linearly changing pressure with arbitrary slope $-\tfrac{1}{2}\beta$, $\beta >0$, as
  \begin{equation} \label{3.2}
\prod_{j=0}^{N-2}\mathrm{d}^2\alpha_j \mathrm{d}\phi \prod_{j=0}^{N-2}(\rho_j^2)^{-1} (\rho_j^2)^{\beta(N-1-j)/2}.
\end{equation}
Surprisingly, relative to this measure the joint distribution of eigenvalues  of $C_N^\diamond $ can be computed in a concise way \cite{KN04}.
We define the Vandermonde determinant
\begin{equation} \label{3.3}
\Delta(z_1,...,z_N) = \prod_{1 \leq i < j \leq N} (z_j - z_i).
\end{equation}
Denoting the eigenvalues of $C_N^\diamond $ by $\mathrm{e}^{\mathrm{i}\vartheta_1},\dots,\mathrm{e}^{\mathrm{i}\vartheta_N}$, their joint (unnormalized) distribution under the measure in \eqref{3.2}
is given by 
\begin{equation} \label{3.4}
\zeta_N^\diamond(\beta) |\Delta(\mathrm{e}^{\mathrm{i}\vartheta_1},..., \mathrm{e}^{\mathrm{i}\vartheta_N})|^\beta \prod_{j=1}^N\mathrm{d}\vartheta_j, \qquad \zeta_N^\diamond(\beta) = 2^{(1-N)} \frac{1}{N!}
\frac{\Gamma(\beta/2)^N}{\Gamma(N\beta/2)},
\end{equation}

Since $\beta$ is a free parameter, one can choose specifically
\begin{equation} \label{3.5}
\beta = \frac{2P}{N}.
\end{equation}
Now the ramp has slope $-P/N$ and, in the limit $N \to \infty$,  close to the lattice point $(1-u)N$, $0 < u <1$, the measure of \eqref{3.2}  will converge to the product measure of \eqref{2.22} with pressure $uP$. Since
\begin{equation} \label{3.6}
\mathrm{tr}\big[\hat{V}(C_N^\diamond)] = \sum_{j=1}^N V(\vartheta_j),
\end{equation}
the Boltzmann weight can be naturally included in \eqref{3.2}. Hence the partition function of the system with boundary conditions is defined by
\begin{eqnarray} \label{3.7}
&&\hspace{-40pt} Z_N^\diamond(P,V) = \int_{[0,2\pi]^{N-1}}\prod_{j=0}^{N-2}\mathrm{d}^2\alpha_j\int_0^{2\pi} \mathrm{d}\phi \prod_{j=0}^{N-2}(\rho_j^2)^{-1} (\rho_j^2)^{P(N-1-j)/N} \exp\big(-\mathrm{tr}[V(C_N^\diamond)]\big) \nonumber\\
&&\hspace{-20pt} =\zeta_N^\diamond(P) \int_{0}^{2\pi} \!\!\mathrm{d}\vartheta_1\dots  \int_{0}^{2\pi}\!\! \mathrm{d}\vartheta_N
\exp\Big( - \sum_{j=1}^N V(\vartheta_j) + P \frac{1}{N} \sum_{j,\ell=1,j\neq\ell}^N  \log|\mathrm{e}^{\mathrm{i}\vartheta_\ell} - \mathrm{e}^{\mathrm{i}\vartheta_j} |\Big).
\end{eqnarray}
In statistical mechanics the probability distribution 
\begin{equation} \label{3.8}
(Z_{\log, N})^{-1}\exp\Big( - \sum_{j=1}^N V(\vartheta_j) + P \frac{1}{N} \sum_{j,\ell=1,j\neq\ell}^N  \log|\mathrm{e}^{\mathrm{i}\vartheta_\ell} - \mathrm{e}^{\mathrm{i}\vartheta_j} |\Big)
\end{equation}
is known as CUE or circular log-gas \cite{F10}. Since the coupling strength is proportional to $1/N$, it is the much studied mean-field version of the log-gas, see \cite{TT20,HL21} and references therein. In the statistical mechanics interpretation $\beta$ is the inverse temperature
and the regime defined through \eqref{3.5} can be viewed as high temperature.
 
 The AL model with boundary conditions has the free energy per site defined through 
\begin{equation} \label{3.9}
F^\diamond\hspace{-1pt}(P,V) = \lim_{N \to \infty} - \frac{1}{N} \log Z_{N}^{\diamond}(P,V).
\end{equation}
Because the pressure ramp has slope $1/N$, in the limit the free energies merely add up as
\begin{equation} \label{3.10}
F^\diamond\hspace{-1pt}(P,V) = \int_0^1 \mathrm{d}u F_\mathrm{AL}(uP,V).
\end{equation}

Before studying the infinite volume free energy, we remark that the CMV matrix $C_N^\diamond$ is still linked to a suitably modified AL dynamics governed by the 
hamiltonian
\begin{equation} \label{3.11}
H_N^\diamond=  \mathrm{tr}\big[C_N^\diamond + {C_N^\diamond}\hspace{-1pt}^*\big].
\end{equation}
Working out the Poisson brackets leads to the evolution equation
\begin{equation} \label{3.12}
\frac{d}{dt}\alpha_j  = \mathrm{i} \rho_j^2 (\alpha_{j-1} + \alpha_{j+1}),\quad \rho_j^2 = (1 - |\alpha_j|^2),
\end{equation}
$j = 0,\dots, N-2$, with the boundary conditions $\alpha_{-1} = -1$ and $\alpha_{N-1} = \mathrm{e}^{\mathrm{i}\phi}$. As before 
$\mathrm{tr}\big[{(C_N^\diamond})^n\big]$ is preserved under the dynamics. However, the a priori measure \eqref{3.2} is no longer stationary.  
The long time dynamics with tied down boundary conditions  differs qualitatively from the one on the ring.

The prefactor in \eqref{3.7} can be easily handled with the result
\begin{equation} \label{3.13}
 \lim_{N \to \infty} - \frac{1}{N} \log\zeta_{N}^{\diamond}(2P/N) = \log(2P).
\end{equation}
For the main term of \eqref{3.7}, one notes that the exponential can be written in terms of the empirical density 
\begin{equation} \label{3.14}
 \varrho_N (w) = \frac{1}{N} \sum_{j = 1}^N \delta(w - \vartheta_j).
\end{equation}
The term reflecting the confining potential $V$ is linear in $\varrho_N$, while the interaction term is quadratic up to the diagonal contribution. 
Thus the limiting free energy 
is determined by a variational principle. We first define the mean-field free energy functional
\begin{equation} \label{3.15}
\mathcal{F}^\mathrm{MF}(\varrho) =  \int _0^{2\pi}\hspace{-6pt}\mathrm{d}w \varrho(w) V(w)      - P\int _0^{2\pi}\hspace{-6pt}\mathrm{d}w\int_0^{2\pi}\hspace{-6pt}\mathrm{d}w'   
\log|\mathrm{e}^{\mathrm{i}w}-\mathrm{e}^{\mathrm{i}w'}|\varrho(w) \varrho(w') + 
\int_0^{2\pi}\hspace{-4pt}\mathrm{d}w \varrho(w) \log \varrho(w).
\end{equation} 
This functional has to be varied over all densities $\varrho$, with the constraints $\varrho(w) \geq 0$ and $\langle  \varrho \rangle = 1$. The minimizer is known to be unique \cite{ST97} and will be denoted by $\varrho^\star$. One then arrives at 
\begin{equation} \label{3.16}
 F^\diamond\hspace{-1pt}(P,V) =  \log(2P) + \mathcal{F}^\mathrm{MF}(\varrho^\star) 
\end{equation}
and hence, using \eqref{3.10},
\begin{equation}\label{3.17} 
F_\mathrm{AL}(P,V) = \partial_P(P\mathcal{F}^\mathrm{MF}(\varrho^\star)) +\log (2P) +1.
\end{equation}

It turns out to be more convenient to absorb $P$ into $\varrho$ by setting $\rho = P\varrho$. Then 
$P \mathcal{F}^\mathrm{MF}(P^{-1}\rho)=  \mathcal{F}(\rho) -P\log P$ with 
the transformed free energy functional
\begin{equation}\label{3.18}
\mathcal{F}(\rho) =  \int _0^{2\pi}\hspace{-6pt}\mathrm{d}w \rho(w) V(w)  - \int _0^{2\pi}\hspace{-6pt}\mathrm{d}w\int_0^{2\pi}\hspace{-6pt}\mathrm{d}w'   
\log|\mathrm{e}^{\mathrm{i}w}-\mathrm{e}^{\mathrm{i}w'}|\rho(w) \rho(w') + 
\int_0^{2\pi}\hspace{-6pt}\mathrm{d}w \rho(w) \log \rho(w).
\end{equation} 
$\mathcal{F}$ has to be minimized under the constraint
\begin{equation}\label{3.19}
\rho(w) \geq 0,\quad \int_0^{2\pi}\hspace{-6pt} \mathrm{d}w\rho(w) =P 
\end{equation}
with minimizer denoted by $\rho^\star$. Then
 \begin{equation}\label{3.20} 
 F_\mathrm{AL}(P,V) =  \partial_P \mathcal{F}(\rho^\star) +\log 2.
 \end{equation}
 
The constraint \eqref{3.19} is removed by introducing the Lagrange multiplier $\mu$ as
  \begin{equation}\label{3.21} 
 \mathcal{F}_\mu(\rho) =  \mathcal{F}(\rho) - \mu \int_0^{2\pi}\hspace{-6pt}\mathrm{d}w \rho(w).
 \end{equation}
A minimizer of  $\mathcal{F}_\mu(\rho)$ is denoted by  $\rho_\mu$ and determined as solution of the Euler-Lagrange equation
\begin{equation}\label{3.22} 
  V(w)   - \mu -  2 \int_0^{2\pi} \hspace{-6pt}\mathrm{d}w'  \log|\mathrm{e}^{\mathrm{i}w}-\mathrm{e}^{\mathrm{i}w'}| \rho_\mu(w') +\log \rho_\mu(w) = 0.
 \end{equation}
The Lagrange parameter $\mu$ has to be adjusted such that
\begin{equation}\label{3.23} 
 P =  \int _0^{2\pi}\hspace{-6pt}\mathrm{d}w  \rho_\mu(w).
 \end{equation}
 To obtain the Ablowitz-Ladik free energy, we differentiate as
 \begin{eqnarray}\label{3.24} 
 && \hspace{-40pt} \partial_P \mathcal{F}(\rho^\star) =
   \int _0^{2\pi} \hspace{-6pt}\mathrm{d}w \partial_P\rho^\star(w) V(w)      - 2  \int _0^{2\pi}\hspace{-6pt} \mathrm{d}w  \int _0^{2\pi} \hspace{-6pt}\mathrm{d}w' \log|\mathrm{e}^{\mathrm{i}w}-\mathrm{e}^{\mathrm{i}w'}| \rho^\star(w)\partial_P\rho^\star(w')\nonumber\\
 && \hspace{40pt} 
 + 1 +  \int _0^{2\pi}\hspace{-6pt} \mathrm{d}w( \partial_P\rho^\star(w) ) \log \rho^\star(w).
  \end{eqnarray}
 Integrating \eqref{3.21} against  $\partial_P\rho^\star$ one arrives at
 \begin{equation}\label{3.25} 
 \partial_P \mathcal{F}(\rho_\mu) = \mu +\log 2
\end{equation}
 and thus 
 \begin{equation}\label{3.26} 
  F_\mathrm{AL}(P,V) =  \mu(P,V)+ \log 2.
 \end{equation}
Sharing with other many-body integrable models, the Ablowitz-Ladik lattice has the property that its free energy 
is determined through an explicit variational problem.\\\\
\textit{Added note}. Independently, T. Grava and G.  Mazzuca \cite{GM21}  discovered the construction just presented.
In case of a confining potential specified by a finite polynomial, for the Ablowitz-Ladik model they prove 
the existence of the infinite volume limit DOS and its connection to the minimizer of the generalized free energy as in
\eqref{4.6}. For thermal equilibrium, they show that the minimizer is a solution of the double
confluent Heun equation.
\section{Density of states}
\label{sec4}
\setcounter{equation}{0}
For the derivation of the hydrodynamic equations,  the GGE average of the conserved fields, $\langle Q^{[n,\sigma]}_0\rangle_{P,V}$, 
 is required for which purpose there are two equivalent methods.
One can start from the microscopic definition and use that  $Q^{[n,\sigma],N}$ depends only on the eigenvalues of the CMV matrix. The other method, employed here, is to simply differentiate the free energy per site. We start with $n=0$ and note that the average log intensity
\begin{equation}\label{4.1} 
\nu= \langle Q^{[0]}_{0}\rangle_{P,V} = \partial_P F_\mathrm{AL}(P,V) = \partial_P\mu(P,V) = \Big(\int _0^{2\pi}\hspace{-6pt} \mathrm{d}w  \partial_\mu\rho_\mu(w) \Big)^{-1},
\end{equation}
where the last equality results from differentiating Eq. \eqref{3.19} as $1 = (\int \partial_\mu\rho_\mu) \mu'(P)$. For $n\geq 1$ we perturb $V$ as $V_\kappa(w) = V(w) + \kappa \varsigma_{n\sigma}(w)$ and differentiate the free energy at $\kappa = 0$.  Then 
\begin{equation}\label{4.2} 
 \langle Q^{[n,\sigma]}_{0}\rangle_{P,V} = \partial_\kappa F_\mathrm{AL}(P,V_\kappa)\big|_{\kappa = 0} = \partial_P \partial_\kappa \mathcal{F}(\rho^\star(P,V_\kappa))\big|_{\kappa = 0} 
 \end{equation}
 and, first introducing the linearization of $\rho^\star$ as
\begin{equation}\label{4.3} 
\partial_\kappa \rho^\star(P,V_\kappa )\big|_{\kappa = 0} =  \rho{^\star}',
\end{equation}
one obtains
\begin{eqnarray}\label{4.4} 
 && \hspace{-40pt}  \partial_\kappa \mathcal{F}(\rho^\star(P,V_\kappa))\big|_{\kappa = 0} =
\int _0^{2\pi}\hspace{-6pt} \mathrm{d}w  \rho^\star(w,P,V)\varsigma_{n\sigma} + \int _0^{2\pi}\hspace{-6pt} \mathrm{d}w V(w) \rho{^\star}'(w) \\
 &&\hspace{-10pt} 
  - 2 \int _0^{2\pi}\hspace{-6pt} \mathrm{d}w \int _0^{2\pi}\hspace{-6pt} \mathrm{d}w'  \log|\mathrm{e}^{\mathrm{i}w}-\mathrm{e}^{\mathrm{i}w'}| \rho{^\star}'(w)\rho^*(w',P,V) + \int _0^{2\pi}\hspace{-6pt} \mathrm{d}w 
  \rho{^*}'(w) \log \rho^\star(w,P,V),\nonumber
  \end{eqnarray}
using that $  \int _0^{2\pi}\hspace{-4pt}\mathrm{d}w \rho{^\star}'(w)= 0$.
 Integrating the Euler-Lagrange equation  \eqref{3.22} at $\mu = \mu(P)$ against $\rho{^\star}'$, the terms on the right side of \eqref{4.4} vanish and
 \begin{equation}\label{4.5} 
 \langle Q^{[n,\sigma]}_{0}\rangle_{P,V} 
 = \int _0^{2\pi}\hspace{-6pt}\mathrm{d}w \big(\partial_P\rho^\star(w,P,V)\big) \varsigma_{n\sigma}(w). 
\end{equation}
Thus in the limit  $N \to \infty$, density of states is given by
\begin{equation}\label{4.6} 
 \rho_\mathrm{Q}(w) = \partial_P\rho^\star(w).
\end{equation}
Naively one might have guessed that the DOS equals $\varrho^\star$. But the linear pressure ramp in the Killip-Nenciu identity amounts to a slightly deviating  result. 

In the literature an Euler-Lagrange equation of the type  \eqref{3.22} is written differently by formally introducing a Boltzmann weight through
 \begin{equation}\label{4.7} 
 \rho_\mu(w) = \mathrm{e}^{-\varepsilon(w)}
\end{equation}
with quasi-energy $\varepsilon(w)$. Then
\begin{equation}\label{4.8} 
 \varepsilon(w) = V(w) -\mu -  2  \int _0^{2\pi}\hspace{-6pt} \mathrm{d}w'   \mathrm{d}w'    \log|\sin(\tfrac{1}{2}(w - w'))|\mathrm{e}^{-\varepsilon(w')}.
 \end{equation}
 In fact,  comparing with \eqref{2.25}, one could absorb $\mu$ into $V$. In general one has to obtain the solution numerically.
 As discussed in \cite{BCS19}, the most efficient method appears to use the nonlinear Fokker-Planck equation related to Dyson's Brownian motion on the circle. If $V=0$, the solution is uniform on the interval $[0,2\pi]$.
 
 We pause for a while to collect a number standard identities, together with their notations.
 The Hilbert space of square integrable functions on $[0,2\pi]$ is denoted by $L^2([0,2\pi], \mathrm{d}w)$ with scalar product
 \begin{equation}\label{4.9} 
\langle f,g\rangle = \int _0^{2\pi}\hspace{-6pt} \mathrm{d}w \bar{f}(w) g(w).
 \end{equation}
 There will be many integrals 
 over $[0,2\pi]$  and a convenient  shorthand is 
  \begin{equation}\label{4.10} 
\langle f \rangle = 
 \langle 1,f\rangle  =  \int _0^{2\pi}\hspace{-6pt} \mathrm{d}w f(w).
 \end{equation}
 Let us define the integral operator 
\begin{equation}\label{4.11}
T\psi(w) = 2 \int _0^{2\pi}\hspace{-6pt} \mathrm{d}w'   \log|\sin(\tfrac{1}{2}(w - w'))|\psi(w'),\quad w \in [0,2\pi].
\end{equation}
Then the Euler-Lagrange equation \eqref{4.8} can be rewritten as 
\begin{equation}\label{4.12} 
\varepsilon (w)  = V(w)  -  \mu  - (T \mathrm{e}^{-\varepsilon})(w).
 \end{equation}
One introduces the  dressing of a function $\psi$  through
\begin{equation}\label{4.13} 
\psi^\mathrm{dr} = \psi + T \rho_\mu \psi^\mathrm{dr},\quad \psi^\mathrm{dr} = \big(1 - T\rho_\mu\big)^{-1} \psi,
\end{equation}
where $\rho_\mu$ is regarded as multiplication operator, i.e. $(\rho_\mu\psi)(w) = \rho_\mu(w)\psi(w) $.
With our improved notation, the DOS in \eqref{4.6} can be written as 
\begin{equation}\label{4.14} 
\rho_{\mathrm{Q}} =  \partial_P\rho_\mu = ( \partial_P\mu)\partial_\mu \rho_\mu = \nu \rho_\mathsf{p}, \quad \partial_\mu \rho_\mu = \rho_\mathsf{p},  \quad \nu\langle\rho_\mathsf{p}\rangle = 1.
 \end{equation}
Differentiating \eqref{4.13} with respect to $\mu$ we conclude
\begin{equation}\label{4.15} 
\rho_\mathsf{p}= (1 - \rho_\mu T)^{-1} \rho_\mu = \rho_\mu(1 - T\rho_\mu)^{-1}\varsigma_0 = \rho_\mu \varsigma_0^\mathrm{dr}.
 \end{equation}
For later purposes, we also state
 \begin{equation}\label{4.16} 
q_{n\sigma} =   \langle Q^{[n,\sigma]}_{0}\rangle_{P,V}  = \nu\langle\rho_\mathsf{p}\varsigma_{n\sigma}\rangle.
\end{equation}
Of physical relevance are $\nu$ and $\nu\rho_\mathsf{p}$, since they encode the GGE average of the conserved fields.
 \section{Average currents, hydrodynamic equations}
\label{sec5}
\setcounter{equation}{0}
 \textit{GGE averaged currents}. Returning to a ring of $N$ sites with periodic boundary conditions, for the conserved field with index $[n,\sigma]$  
 the current from $j-1$ to $j$ is denoted by $J_{j}^{[n,\sigma],N}$ and  the current from $j$ to $j+1$ by $J_{j+1}^{[n,\sigma],N}$.
Then the conserved fields satisfy a continuity equation of the form
  \begin{equation}\label{5.1} 
\frac{d}{dt} Q_j^{[n,\sigma],N} = J_{j}^{[n,\sigma],N} - J_{j+1}^{[n,\sigma],N} = \{Q_j^{[n,\sigma],N},H_N\}, \qquad J_{j}^{[n],N} =
J_{j}^{[n,+],N} + \mathrm{i} J_{j}^{[n,-],N}.
\end{equation}
 As explained in more detail in Appendix A, the current densities are local and shift invariant in the sense that 
 \begin{equation} \label{5.2}
J^{[n],N}_{j+1}(\alpha) = J^{[n],N}_j(\tau \alpha).
\end{equation}
 The lowest index current densities are  
  \begin{equation}
  \label{5.3} 
 J_{j}^{[0],N} = 2Q_j^{[1,-],N}, \quad J_{j}^{[1],N} = \mathrm{i}\big(-\rho_{j-1}^2\alpha_{j-2} \bar{\alpha}_{j} +  |\alpha_{j-1}|^2\big). 
 \end{equation}
For higher currents one has to rely on an  abstract argument, see Appendix A.
Actually such expressions are not so helpful when trying to compute the GGE averaged currents.
Fortunately there is a generic argument \cite{S21,S20}, which applies also to the AL system.

We start with the fields and define the infinite volume correlator 
\begin{equation}\label{5.4} 
C_{m\sigma,n\sigma'}(j-i) = \langle Q_j^{[m,\sigma]}Q_i^{[n,\sigma']} \rangle_{P,V}^\mathrm{c},
\end{equation}
where the superscript $^\mathrm{c}$ denotes truncation, or connected correlation, $\langle gf \rangle^\mathrm{c} = \langle gf \rangle -\langle g\rangle\langle f \rangle$. Truncated correlations decay rapidly to zero and the field-field susceptibility matrix is given by
\begin{equation}\label{5.5} 
C_{m\sigma,n\sigma'} = \sum_{j\in \mathbb{Z}}C_{m\sigma,n\sigma'}(j) = \langle Q^{[m,\sigma]};Q^{[n,\sigma']}\rangle_{P,V}, 
\end{equation}
where the right hand is merely a convenient notation for the sum. $C_{m\sigma,n\sigma'}$ is the matrix of second derivatives of the generalized free energy.
Correspondingly we introduce the field-current correlator
\begin{equation}\label{5.6} 
B_{m\sigma,n\sigma'}(j-i) = \langle J_j^{[m,\sigma]} Q_i^{[n,\sigma']} \rangle_{P,V}^\mathrm{c}, \quad B_{m\sigma,n\sigma'} =  \sum_{j\in \mathbb{Z}} B_{m\sigma,n\sigma'}(j).
\end{equation}
Despite its apparently asymmetric definition, $B$ satisfies
\begin{equation}\label{5.7} 
B_{m\sigma,n\sigma'}(j) = B_{n\sigma',m\sigma}(-j).
\end{equation}
To prove, we employ the conservation law and spacetime stationarity to arrive at
\begin{eqnarray}\label{5.8} 
&&\hspace{-70pt}\partial_j \langle J_j^{[m,\sigma]}(t) Q_0^{[n,\sigma']}(0) \rangle_{P,V}^\mathrm{c} 
= -\partial_t  \langle Q_j^{[m,\sigma]}(t) Q_0^{[n,\sigma']}(0) \rangle_{P,V}^\mathrm{c}\nonumber\\[0.5ex]
&&\hspace{37pt} = -\partial_t \langle Q_0^{[m,\sigma]}(0) Q_{-j}^{[n,\sigma']}(-t) \rangle_{P,V}^\mathrm{c} 
= \partial_j \langle Q_0^{[m,\sigma]}(0) J_{-j}^{[n,\sigma']}(-t) \rangle_{P,V}^\mathrm{c},
\end{eqnarray}
denoting the difference operator by $\partial_jf(j) = f(j+1) - f(j)$.
Setting $t=0$, the difference $\langle J_j^{[m,\sigma]} Q_0^{[n,\sigma']} \rangle_{P,V}^\mathrm{c} -  \langle J_{-j}^{[n,\sigma']} Q_0^{[m,\sigma]} \rangle_{P,V}^\mathrm{c}$ is constant in $j$. Since truncated correlations decay to zero,
this constant has to vanish, which yields \eqref{5.7}. In particular, the field-current susceptibility matrix is symmetric,
\begin{equation}\label{5.9} 
B_{m\sigma,n\sigma'} = B_{n\sigma',m\sigma}.
\end{equation}

Using  this symmetry, we consider the $P$-derivative of the average current, 
 \begin{equation}\label{5.10}
   \partial_P \langle J^{[n,\sigma]}_0 \rangle_{P,V} = - B_{n\sigma,0} = -  B_{0,n\sigma}
   = - 2\langle Q^{[1,-]};Q^{[n,\sigma]}\rangle_{P,V},
\end{equation}
since $J^{[0]} = 2Q^{[1,-]}$ by \eqref{5.3}. We easily arrived at a very surprising identity. The $P$-derivative of the average current equals  a particular covariance of the eigenvalue fluctuations. In \cite{S21}, Section 4, the fluctuations of eigenvalues for the GUE mean-field log-gas are handled. To switch from GUE to CUE in essence amounts to  a notational change. The joint distribution of eigenvalues is stated in \eqref{3.8}. Their asymptotic density is
$\varrho^\star$, as minimizer of the variational problem \eqref{3.15}.  The corresponding fluctuation field  is defined through 
\begin{equation}\label{5.11}
\phi_N(f) = \frac{1}{\sqrt{N}} \sum_{j=1}^N \big(f(\vartheta_j) - \langle\rho^\star f \rangle\big) = \int_0^{2\pi} \hspace{-6pt}\mathrm{d}w f(w) \phi_N(w)
\end{equation} 
with $f$ some smooth test function on the circle $[0,2\pi]$. As $N\to \infty$, $\phi_N$ converges to a Gaussian field with covariance 
\begin{equation}\label{5.12}
\langle \tilde{f}, C^\sharp f \rangle = 
\langle  \tilde{f},(1 - P \varrho^\star T)^{-1} \varrho^\star f\rangle - \nu P 
\langle \tilde{f},(1 - P \varrho^\star T)^{-1}\varrho^\star \rangle \langle (1 - P \varrho^\star T)^{-1}\varrho^\star, f\rangle,
\end{equation} 
where $\varrho^\star$ is regarded as multiplication operator. The subtraction arises because the number of eigenvalues does not fluctuate.

As in the case of the free energy, since the pressure is varying as $1/N$, the fluctuation covariance is adding up, resulting in
\begin{equation}\label{5.13}
\int_0^1 \mathrm{d}u \langle Q^{[1,-]};Q^{[n,\sigma]} \rangle_{uP,V} = \langle \varsigma_{1-}, C^\sharp \varsigma_{n\sigma}\rangle.
\end{equation}
Therefore, using identity \eqref{5.10}, one arrives at
 \begin{equation}\label{5.14}
\partial_P\big(\langle J^{[n,\sigma]}_0 \rangle_{P,V} + 2P\langle \varsigma_{1-}, C^\sharp \varsigma_{n\sigma}\rangle\big) = 0,
\end{equation}
implying that the round bracket has to be independent of $P$, in particular equal to its value at $P=0$.  Since $\langle \varsigma_{1-}, C^\sharp \varsigma_{n\sigma}\rangle$ is bounded in $P$, the second summand vanishes at $P=0$. For the first summand, one notes that 
in the limit $P \to 0$, for each $j$ the a priori measure \eqref{2.22} becomes uniform  on the unit circle and the CMV matrix turns diagonal, since $\rho_j^2 \to 0$. Denoting 
$\alpha_j = \mathrm{e}^{\mathrm{i}\phi_j}$, $\phi_j \in [0,2\pi]$, in the limit $P \to 0$ the GGE (2.19) converges  to
\begin{equation} \label{5.15}
(Z_N)^{-1}\prod_{j=0}^{N-1}\mathrm{d}\phi_j \exp\Big( - \sum_{j=0}^{N-1} V(\phi_{j+1} - \phi_{j})\Big),\qquad \phi_N = \phi_0 
\end{equation}
Using \eqref{5.3}, one observes that $\langle J_0^{[1]}  \rangle_{0,V} =\mathrm{i}$. By a direct computation  $\langle J_0^{[2]}  \rangle_{0,V} = 0$. To extend the average to general $n$ seems to be difficult, since a sufficiently explicit formula for $J_0^{[n]}$ is missing.
We assume that $\langle J_0^{[n]}  \rangle_{0,V} = d_n$ with some constant $d_n$ independent of $V$.
  Next we substitute $P\varrho^* = \rho_\mu$ with the result
\begin{equation}\label{5.16}
P\langle \varsigma_{1-}, C^\sharp \varsigma_{n\sigma}\rangle    = \langle \varsigma_{1-}, (1 -  \rho_\mu T)^{-1}\rho_\mu \varsigma_{n\sigma}\rangle
    -    \nu \langle \varsigma_{1-}, (1 -  \rho_\mu T)^{-1} \rho_\mu\rangle  \langle \varsigma_{n\sigma}, (1 -  \rho_\mu T)^{-1} \rho_\mu\rangle.
\end{equation}
Noting that 
$(1- \rho_\mu T)^{-1} \rho_\mu$ is a symmetric operator, one  finally arrives at
\begin{equation}\label{5.17}
 \langle J^{[n,\sigma]}_0 \rangle_{P,V} -d_n =  -2 \big(\langle \rho_\mu \varsigma_{1-}^\mathrm{dr}   \varsigma_{n\sigma}\rangle -  q_{1-}\langle \rho_\mathsf{p}\varsigma_{n\sigma}\rangle\big) 
 ,   \qquad q_{1-} = \nu \langle \rho_\mathsf{p}\varsigma_{1-}\rangle.
\end{equation}

In the conventional scheme of generalized hydrodynamics, the average currents are written somewhat differently. Firstly by linearity, there is some function
 $\rho_J(w)$ on $[0,2\pi]$ such that 
 \begin{equation}\label{5.18}
  \langle J^{[n,\sigma]}_0 \rangle_{P,V}- d_n = \langle \varsigma_{n\sigma}\rho_J\rangle.
\end{equation}
For the currents, $\rho_J$ plays the same role as $\rho_Q$ for the conserved fields. However $\rho_J$ cannot have a definite sign.
 Secondly one writes 
  \begin{equation}\label{5.19}
\rho_J = -2\rho_\mathrm{p} \big(v^\mathrm{eff} - q_{1-}\big), \qquad q_{1-} = \langle Q^{1,-}_0\rangle_{P,V},
\end{equation}
 with $v^\mathrm{eff}$ the effective velocity. The effective velocity can be written more concisely as
 \begin{equation}\label{5.20} 
v^\mathrm{eff} = \frac{\varsigma_{1-}^\mathrm{dr}}{\varsigma_0^\mathrm{dr}}\,, 
 \end{equation}
see \cite{S21}, Section 6.\\\\
\textit{Hydrodynamic equations.} On the hydrodynamic scale the local GGE is characterized by the log intensity $\nu$ and the CMV density of states $\nu\rho_\mathsf{p}$, both of which now become spacetime dependent. Merely inserting the average currents, 
and since $d_n$ has been assumed to be a constant, one arrives at the Euler type hydrodynamic evolution equations,
\begin{eqnarray}\label{5.21} 
&&\partial_t \nu(x,t) + 2\partial_x q_{1-}(x,t) = 0,\nonumber\\[0.5ex]
&& \partial_t\big(\nu(x,t) \rho_\mathsf{p}(x,t;v)\big) - 2\partial_x\big((v^\mathrm{eff}(x,t;v) - q_{1-}(x,t))\rho_\mathsf{p}(x,t;v)\big) = 0.
\end{eqnarray}
This equation is based on the assumption of local GGE. To actually establish such an equation from the underlying AL model
seems to be a difficult task.
 
 As a most remarkable feature of generalized hydrodynamics,  the equations can be  transformed explicitly to a quasilinear 
form, see \cite{S21}, Section 6. 
For this purpose we rewrite the identity  in \eqref{4.15} as 
\begin{equation}\label{5.22} 
\rho_\mu =  \rho_\mathsf{p}(1+ (T\rho_\mathsf{p}))^{-1},
\end{equation}
now regarded as the nonlinear mapping $\rho_\mathsf{p} \mapsto \rho_\mu$. Then
Eq. \eqref{5.21} assumes the normal form
\begin{equation}\label{5.23} 
 \partial_t \rho_\mu - 2\nu^{-1}(v^\mathrm{eff} - q_{1-})\partial_x \rho_\mu = 0.
\end{equation}
Thus the linearization operator is in fact merely multiplication by $-2\nu^{-1}(v^\mathrm{eff} - q_{1-})$, in other words the operator is diagonal. 

In \eqref{2.1} we followed a standard convention, which amounts to the free dispersion relation $E(p) = 2(1- \cos p)$. Upon adopting 
$E(p) = 1- \cos p$ the extra factors of 2 in \eqref{5.21} and \eqref{5.23} would be removed. 
 \section{Modified Korteweg-de Vries equation}
\label{sec6}
\setcounter{equation}{0}
Instead of the hamiltonian $H_N$ of \eqref{2.5}, one can choose
\begin{equation}
\label{6.1}
\breve{H}_N = - \mathrm{i}\sum_{j=0}^{N-1} \big( \alpha_{j-1} \bar{\alpha}_{j} - \bar{\alpha}_{j-1} \alpha_{j}\big) = - \mathrm{i} \,\mathrm{tr}\big[C_N - C_N^*\big].
\end{equation}
Then
\begin{equation} \label{6.2}
\frac{d}{dt}\alpha_j = \{\alpha_j,\breve{H}_N\}_\mathrm{AL}   =  \rho_j^2 (\alpha_{j+1} - \alpha_{j-1}) ,
\end{equation} 
which is known as Schur  flow \cite{G06}.
Through a formal Taylor expansion, in \cite{AL76}  it is argued that the continuum limit of Eq. \eqref{6.2} yields the modified 
Korteweg-de Vries equation
\begin{equation} \label{6.3}
\partial_t u= \partial_x^3u - 6 u^2\partial_x u,
\end{equation} 
which is a good reason to briefly touch upon \eqref{6.2}.

As before $\alpha_j  \in \mathbb{D}$ and the conservation laws remain  unchanged. However the currents have to be modified from $J^{[n]}$ to $\breve{J}^{[n]}$.
For the log intensity current one finds
\begin{equation} \label{6.4}
\breve{J}^{[0],N}_j =  2Q^{[1,+],N}_j= H_{N,j}.
\end{equation} 
 The arguments of Section 5 can be repeated, ad verbatim. In the hydrodynamic equations \eqref {5.21}, $q_{1-}$ is replaced by $q_{1+}$
and the effective velocity turns to 
 \begin{equation}\label{6.5} 
v^\mathrm{eff} = \frac{\varsigma_{1+}^\mathrm{dr}}{\varsigma_0^\mathrm{dr}}\,. 
 \end{equation}
Such interchange of the roles of momentum and energy is already familiar from the relativistic sinh-Gordon  quantum field theory \cite{CDY16}. 

Such a discussion misses however an interesting point. The wave field of the modified Korteweg-de Vries equation is real-valued, a feature which is maintained in the discrete approximation. 
If in \eqref{6.2} one chooses initially a real field $\alpha$, then it stays real throughout time. From the perspective of GGE, such initial conditions  amount 
to a set of measure zero and one has to reconsider the analysis.  
Fortunately the relevant transformation formula is already proved in \cite{KN04}. 
To avoid duplication of symbols, in the remainder of this section $\alpha_j \in \mathbb{R}$, hence $\bar{\alpha}_j = \alpha_j$ everywhere. The equations of motion read
\begin{equation} \label{6.6}
\frac{d}{dt}\alpha_j = \rho_j^2 (\alpha_{j+1} - \alpha_{j-1}),\qquad \alpha_j \in [-1,1], \qquad \rho_j^2 = 1 -\alpha_j^2.
\end{equation}
While an obvious hamiltonian structure is lost, one readily checks that the a priori measure
\begin{equation} \label{6.7}
\prod_{j=0}^{N-1}\mathrm{d}\alpha_j (\rho_j^2)^{P-1}= \prod_{j=0}^{N-1}\mathrm{d}\alpha_j (\rho_j^2)^{-1}\exp\big(-PQ^{[0],N}\big) 
\end{equation}
is still stationary under the dynamics. As before the densities of the conserved fields are 
\begin{equation} \label{6.8}
Q_j^{[m]} = (C^m)_{j,j},
\end{equation} 
 $m = 1,2,...\,$. In particular, 
\begin{equation} \label{6.9}
Q^{[0]}_j =  -\log\rho_j^2,\quad Q^{[1]}_j = -\alpha_{j-1}\alpha_j,\quad Q^{[2]}_j = \alpha_{j-1}^2\alpha_{j}^2
- \rho_{j-1}^2\alpha_{j-2}\alpha_{j} - \rho_{j}^2\alpha_{j-1}\alpha_{j+1}.
\end{equation}
There is no longer a distinction of $\pm$. For the log intensity current, $J^{[0]}_j = 2Q^{[1]}_j$. Thus  $J^{[0]}$ is conserved and the indirect method for computing the average currents is still in place, see Section 5.

Since now $C_N$ is a real matrix, its eigenvalues come in pairs. If $\mathrm{e}^{\mathrm{i}\vartheta_j}$ is an eigenvalue,  so is $\mathrm{e}^{-\mathrm{i}\vartheta_j}$.
For a system of size $N$, there are only $n =N/2$ independent eigenvalues. Rather than using a DOS reflecting such symmetry, it is more effective to restrict the eigenvalues as $0 \leq \vartheta_j \leq \pi$, subsequently setting $y_j = \cos \vartheta_j$. 
The empirical DOS is given by  
\begin{equation} \label{6.10}
\rho_{Q,n}(w) \mathrm{d} w   = \frac{1}{n} \sum_{j=1}^n \delta (w - y_j) \mathrm{d}w, \quad w \in [-1,1], 
\end{equation}
where the more convenient $n$ as size parameter is used. In the limit $n \to \infty$, $\rho_{Q,n}(w)$ converges to the deterministic limit $\rho_{Q}(w)$.
The GGE expectations are then
\begin{equation} \label{6.11}
\langle Q_0^{[m]}\rangle_{P,V} = 2\int _{-1}^{1}\hspace{-6pt}\mathrm{d}w\rho_{Q}(w) \varsigma_m(w), \qquad  \varsigma_m(w) = \cos(m\vartheta),  \quad w = \cos\vartheta,
\end{equation}
i.e. being the $m$-th Chebyshev polynomial. The confining potential transforms to
\begin{equation} \label{6.12}
V_\mathrm{kdv}(w) = \sum_{m=1}^\infty \mu_m \varsigma_m(w) 
\end{equation}
with real chemical potentials $\mu_m$.

As proved in \cite{KN04}, under the measure 
\begin{equation} \label{6.13}
\prod_{j=0}^{2n-2}(1-\alpha_j^2)^{-1}(1-\alpha_j^2)^{\beta(2n-j-1)/4}(1-\alpha_j)^{a+1-(\beta/4)}(1 + (-1)^j \alpha_j)^{b +1 - (\beta/4)} \mathrm{d} \alpha_j
\end{equation}
on $[-1,1]^{2n-1}$, $\beta >0$, and $a,b > -1 +(\beta/4)$, the joint (unnormalized) distribution of the eigenvalues of $C_N$, imposing $\alpha_{2n-1} = 1$, is given by  
\begin{equation} \label{6.14}
\zeta_n^\diamond(\beta)2^{\kappa} |\Delta(2 y_1,...,2 y_n)|^\beta \prod_{j=1}^{n}(1 - y_j)^a(1 + y_j)^b \mathrm{d}y_j
\end{equation}
on $[-1,1]^n$. One notes $\beta$ times the energy of the repulsive log gas with the single site a priori weight given by the Jacobi polynomial on $[-1,1]$. 
The normalization $\zeta_n^\diamond$ 
is defined in \eqref{3.4} and $\kappa = (n-1)(-\tfrac{1}{2}\beta +a +b +2)$.  
To achieve a  pressure ramp of slope $-P/2n$, one has to set 
\begin{equation} \label{6.15}
\beta = \frac{2P}{n}, \qquad a = b = -1 +\tfrac{1}{4} \beta.
\end{equation}
Thus as before, one has to study the high temperature regime, this time for the $\beta$ Jacobi ensemble. Since $\kappa = 0$  \eqref{6.14} becomes
\begin{equation} \label{6.16}
\zeta_n^\diamond(\beta) |\Delta(2y_1,...,2y_n)|^\beta \prod_{j=1}^{n}(1 - y_j)^{-1 +(\beta/2)}(1 + y_j)^{-1 +(\beta/2)} \mathrm{d}y_j,
\end{equation} 
while  \eqref{6.13} turns to
\begin{equation} \label{6.17}
\prod_{j=0}^{2n-2}(1-\alpha_j^2)^{-1}(1-\alpha_j^2)^{P(2n-j-1)/2n} \mathrm{d} \alpha_j.
\end{equation}

Now the strategy of Section 3 is in force. We add a confining potential. Then the asymptotic DOS is obtained by minimizing the mean-field free energy 
\begin{eqnarray} \label{6.18}
&&\hspace{-56pt}\mathcal{F}^\mathrm{KdV}(\varrho) =  \int _{-1}^{1}\hspace{-6pt}\mathrm{d}w \varrho(w) V_\mathrm{kdv}(w) +   \int _{-1}^{1}\hspace{-6pt}\mathrm{d}w \varrho(w) 
\log(1-w^2)\nonumber\\
&&\hspace{0pt}    - P\int _{-1}^{1}\hspace{-6pt}\mathrm{d}w\int_{-1}^{1}\hspace{-6pt}\mathrm{d}w'   
\log(2|w - w'|)\varrho(w) \varrho(w') + 
\int_{-1}^{1}\hspace{-4pt}\mathrm{d}w \varrho(w) \log \varrho(w).
\end{eqnarray} 
$\mathcal{F}^\mathrm{KdV}$ has to be minimized over all $\rho \geq 0$ with $\int\mathrm{d}w \varrho(w)=1$ and boundary condition 
$\rho(-1) = \rho(1)$.

Actually our mean-field limit is somewhat singular, since at $a=-1 = b$ the a priori measure is not integrable.
The quadratic energy term is repulsive at short distances, but the linear term with $\log(1-w^2)$ pushes the eigenvalues towards the two end points. 
It is not so obvious, whether and how the two terms balance. Fortunately, the particular case $V_\mathrm{kdv} = 0$ has been studied 
in the recent contributions on the $\beta$-Jacobi ensemble \cite{FM21,TT21}.  Studied is the asymptotic DOS \eqref{6.10} with parameters $\beta = 2\alpha/n$, hence $\alpha = P$, while the parameters $a >-1$ and $b >-1$. Obtained is the exact density of states
in terms of the hypergeometric function $_{2}F_{1}$. As communicated by K.D. Trinh, their proof works also for the limiting cases of interest here.  Simply inserting the values $a=-1,b=-1$ in the general formula,  a well-defined DOS is obtained. A confining potential will modify the DOS, but the balance between terms should persist.

Surprisingly, the confining potential $V_\mathrm{kdv}$ is corrected by the $\log(1-w^2)$ potential, which is attractive and  favors the accumulation of eigenvalues near the two boundary points $\pm1$. The prior computation of the average currents is carried out for fixed $V$,
which has now to be corrected to by $V_\mathrm{cor}(w) = V_\mathrm{kdv}(w) + \log(1- w^2)$. $\rho_\mu$ and $\rho_\mathrm{p}$ depend on $V_\mathrm{cor}$.
Also the dressing operator $T$ is changed to
\begin{equation}\label{6.19}
T\psi(w) =  \int _{-1}^{1}\hspace{-6pt} \mathrm{d}w'\log(2|w - w'|)\psi(w') 
,\quad w \in [-1,1].
\end{equation}
With these modifications, the hydrodynamic equations are derived along the standard route. In particular, the effective velocity is still given by
\begin{equation}\label{6.20}
v^\mathrm{eff} = \frac{\varsigma_{1}^\mathrm{dr}}{\varsigma_0^\mathrm{dr}},
\end{equation}
where now the dressing operator $T$ from \eqref{6.19} has to be used. Also $q_1$ is modified to
\begin{equation}\label{6.21}
q_1 
= 2\nu \int _{-1}^{1}\hspace{-2pt} \mathrm{d}w \rho_\mathrm{p}(w) w.
\end{equation}

 \section{Discussion}
 \label{sec7}
\setcounter{equation}{0}
For the defocusing discrete NLS in one dimension, we established the form of the hydrodynamic equations. As a novel feature, their structure is determined 
by the mean-field version of the log gas corresponding to CUE random matrices.   Our analysis is pretty much on the same level as the one for the Toda lattice. Only the handling of average currents in the limit $P\to 0$
is incomplete. We hope to return to this point in the future.

The Toda lattice is linked to the $\beta$-GUE ensemble at small $\beta$, i.e. high temperatures, such that energy and entropy balance. Our results explain how the Ablowitz-Ladik system is connected 
to $\beta$-CUE ensemble at small repulsion. Thus one might wonder whether other classical matrix ensembles are linked to yet to be identified integrable dynamics.
For the discrete modified KdV equation, as an unexpected feature the TBA equations pick up a correction to the confining potential $V_\mathrm{kdv}$.

 The reported results leave me with a puzzle. In generalized hydrodynamics the accepted expression for the effective velocity is
\begin{equation}\label{7.1}
v^\mathrm{eff} = \frac{[E']^\mathrm{dr}}{[p']^\mathrm{dr}}
\end{equation}
with parametrically given dispersion relation $(p,E)$ \cite{CDY16}. For the Toda lattice $E(p) = \tfrac{1}{2}p^2$ and $E'(p) = p$, $p'= 1$.
The kernel defining the dressing transformation is the two-particle scattering shift, which for Toda is $2\log|w - w'|$. So what is the scattering shift for the discretized nonlinear Schr\"{o}dinger equation and how is  the rule \eqref{7.1} connected to  either \eqref{5.20},
or \eqref{6.5}, or \eqref{6.20}?\\\\
 
\appendix
\section{The general Ablowitz-Ladik system}
\label{appA}
\setcounter{equation}{0}
The defocusing AL model is studied in  \cite{N05,N06}. In particular,  it is proved that the CMV matrix $C_N$ determines the local conservation laws. Here we remark that  the algebra in \cite{N05} actually holds at greater generality. 

Already in the original contribution \cite{AKNS74}, see also \cite{ABT04}, it was noted that an 
often more convenient formulation is the coupled system
\begin{eqnarray} \label{A.1}
&&\hspace{1pt}\frac{d}{dt}q_j  = \mathrm{i} \rho_j^2 (q_{j-1} + q_{j+1}),\nonumber\\
&&\hspace{1pt}\frac{d}{dt}r_j  = -\mathrm{i} \rho_j^2 (r_{j-1} + r_{j+1}),\qquad \rho_j^2 = 1 - q_jr_j,
\end{eqnarray}
with $q_j \in \mathbb{C}$, $r_j \in \mathbb{C}$. (Following the quantum convention, and \cite{N06}, for us the Schr\"{o}dinger equation reads
$\mathrm{i}\partial_t \psi = -\partial_x^2 \psi$, which amounts to reversing time when compared to \cite{ABT04}). As before, we consider a ring with $N$ sites,  $N$ even.
For better readability we will drop the index $N$. The defocusing case corresponds to setting $q_j = \alpha_j$, $r_j = \bar{\alpha}_j$, while the physically equally interesting focusing case to $q_j = \alpha_j$, $r_j = -\bar{\alpha}_j$.

We introduce two classes of CMV matrices. Their building blocks are
\begin{equation} \label{A.2}
\Xi_j =
\begin{pmatrix}
r_j & \rho_j \\
\rho_j & - q_j  \\
\end{pmatrix},
\qquad \tilde{\Xi}_j =
\begin{pmatrix}
q_j & \rho_j \\
\rho_j & -r_j  \\
\end{pmatrix}.
\end{equation}
The matrices $L,\tilde{L}$ and $M,\tilde{M}$ are constructed by the same scheme as before and the CMV matrix is defined through
\begin{equation}\label{A.3} 
C = LM,\qquad \tilde{C} = \tilde{L}\tilde{M},
\end{equation}
i.e. the tilde operation amounts to interchanging $q$ and $r$.
The pair $(q_j,r_j)$ is viewed as canonical coordinates. Then the weighted Poisson bracket generalizes to
\begin{equation} \label{A.4}
\{f,g\} =  \mathrm{i} \sum_{j=0}^{N-1} \rho_j^2\Big(\frac{\partial f}{\partial{r_j}} 
\frac{\partial g}{\partial q_j}   - \frac{\partial f}{\partial{q_j}} 
\frac{\partial g}{\partial{r_j}} \Big).
\end{equation}
The time evolution in \eqref{A.1} is generated by the hamiltonian
\begin{equation}\label{A.5} 
H = - \sum_{j=0}^{N-1}\big(q_jr_{j+1} +r_jq_{j+1}\big) =  \mathrm{tr}\big[C + \tilde{C}\big].
\end{equation}
As before the log intensity field,
\begin{equation} \label{A.6} 
Q^{[0]} =  - \sum_{j=0}^{N-1}  \log(\rho_j) = \tilde{Q}^{[0]},
\end{equation}
is conserved. 
The complete tower of conserved fields is of the form
\begin{equation}\label{A.7} 
Q^{[n]} = \mathrm{tr}\big[C^n\big],\qquad \tilde{Q}^{[n]} = \mathrm{tr}\big[\tilde{C}^n\big]
\end{equation}
with densities
\begin{equation}\label{A.8} 
Q^{[n]}_j = (C^n)_{j,j},\qquad \tilde{Q}^{[n]}_j = (\tilde{C}^n)_{j,j}.
\end{equation}
These densities are shift covariant in the sense
\begin{equation}\label{A.9} 
Q^{[n]}_{j+1}(q,r) = Q^{[n]}_j(\tau q,\tau r),\qquad \tilde{Q}^{[n]}_{j+1}(q,r) = \tilde{Q}^{[n]}_j(\tau q,\tau r),
\end{equation}
where $(\tau q)_j = q_{j+1}$ $\mathrm{mod}(N)$. To confirm one introduces the unitary shift matrix $S$ through $
(S^*AS)_{i,j} = A_{i+1,j+1}$ $\mathrm{mod}(N)$. Then, denoting by ${}^\mathrm{T}$ the transpose,
\begin{equation} \label{A.10}
S^\mathrm{T}C({q,r})S =S^\mathrm{T}L(q,r)SS^\mathrm{T}M(q,r) S =  M(\tau q,\tau r) L(\tau q,\tau r) = C(\tau q,\tau r)^\mathrm{T},
\end{equation}
which implies  \eqref{A.9}. 

To extend the  Lax pair relations \eqref{2.11}, one  notes that
$(C_N)_{i,j}(\alpha,\bar{\alpha})$ is a polynomial in $\alpha = (\alpha_0,...,\alpha_{N-1})$ and $\bar{\alpha} = (\bar{\alpha}_0,...,\bar{\alpha}_{N-1})$, which in \cite{N05} are regarded as independent variables. The Poisson bracket generates some
other polynomial and the operation $(\bar{C}_N)_{i,j}$ interchanges the roles of $\alpha,\bar{\alpha}$. Therefore the algebra in \cite{N05} 
persists upon replacing  $\alpha_j$ by $q_j$  and $\bar{\alpha}_j$ by $r_j$ and properly translating the operations ${\hspace{3pt}}^*$ and $\bar{\hspace{3pt}}\,\,$.
Thereby one arrives at 
\begin{equation}\label{A.11}
\{C,\mathrm{tr}(C)\} = \mathrm{i} [C,C_+],\qquad \{C,\mathrm{tr}(\tilde{C})\} = \mathrm{i} [C,(\tilde{C}_+)^\mathrm{T}].
\end{equation}
Since the Poisson bracket acts as a derivative, one deduces
\begin{equation} \label{A.12}
\{C^n,\mathrm{tr}(C)\} = \sum_{m=0}^{n-1} C^m\mathrm{i}[C,C_+]C^{n-m-1} = \mathrm{i} [C^n,C_+], 
\end{equation}
and similarly 
\begin{equation} \label{A.13}
\{C^n,\mathrm{tr}(\tilde{C})\} =  \mathrm{i} [C^n,(\tilde{C}_+)^\mathrm{T}]. 
\end{equation}
As claimed, $Q^{[n]}$ and $\tilde{Q}^{[n]}$ are conserved. In addition,  mutual Poisson brackets vanish,
\begin{equation} \label{A.14}
\{Q^{[n]}, Q^{[n']}\} = 0, \qquad \{Q^{[n]}, \tilde{Q}^{[n']}\} = 0, \qquad \{\tilde{Q}^{[n]}, \tilde{Q}^{[n']}\} = 0\bigskip.
\end{equation}

The densities \eqref{A.8} can still be expanded as a sum over weighted  random walks on a checkerboard. 
 The matrix element $(C^n)_{i,j}$ is then the sum over all $2n$ step walks starting at $i$ and ending
 at $j$. Each walk represents a particular polynomial obtained by taking the product of local weights along the walk. The weights are\bigskip\\
 \begin{tabular}{ll}
 \hspace{30pt}$\rho_j$ &  for the diagonal steps $j \leadsto j+1$ and $j+1 \leadsto j$,\\
 \hspace{30pt}$r_{j}$ & for the horizontal step $j \leadsto j$  in case the lower square is black, \\
 \hspace{30pt}$-q_{j-1}$ & for the horizontal step $j \leadsto j$  in case the upper square is black. \\
 \end{tabular} 
 \bigskip\\
For $\tilde{Q}^{[n]}$ the roles of $q_j$ and $r_j$ are exchanged. 

In  \cite{ABT04}, Chapters 3.2 and 3.4, also conserved fields are discussed and stated is a recursion relation iteratively determining conserved fields. However, these fields are 
nonlocal and one still would have to follow the ``logarithmic  subtraction procedure" to arrive at their local version,
see \cite{S21}, Section 11.

Of particular interest are the current densities. In \eqref{A.14} the matrix $C_+$ has nonvanishing matrix elements only for
$(j,j+\ell)$ with $\ell = 0,1,2$. The terms with $\ell = 0$ cancel and
\begin{eqnarray} \label{A.15}
&&\hspace{-40pt}\{(C^n)_{j,j},H\} \nonumber\\
&&\hspace{-20pt} = \sum_{\ell = 1,2}\mathrm{i}\Big(C_{j-\ell,j}(C^n)_{j,j-\ell}  -C_{j,j+\ell}(C^n)_{j+\ell,j} + \tilde{C}_{j,j+\ell}(C^n)_{j,j+\ell}  - \tilde{C}_{j-\ell,j}(C^n)_{j-\ell,j}\Big).  
\end{eqnarray}
This looks like a shift difference, but it is not, since the off-diagonal matrix elements are only two-periodic.
 For $n=1$ one obtains 
\begin{equation} \label{A.16} 
\{C_{j,j},\mathrm{tr}(C)\} = \mathrm{i}\big(- \rho_{j-1}^2 q_{j-2}r_{j} + \rho_{j}^2 q_{j-1}r_{j+1} \big),
\end{equation}
while 
\begin{equation} \label{A.17} 
\{C_{j,j},\mathrm{tr}(\tilde{C})\} = \mathrm{i}\big(\rho_{j}^2 q_{j+1}r_{j+1} - \rho_{j-1}^2 q_{j-2}r_{j-2} + \rho_{j}^2\rho_{j+1}^2 - 
 \rho_{j-2}^2 \rho_{j-1}^2\big)=  \mathrm{i}\big( q_{j-1}r_{j-1} - q_{j}r_{j}\big)
\end{equation}
for even $j$ and
\begin{equation} \label{A.18} 
\{C_{j,j},\mathrm{tr}(\tilde{C})\} = \mathrm{i}\big(\rho_{j}^2 q_{j-1}r_{j-1} - \rho_{j-1}^2 q_{j}r_{j}  \big)=  \mathrm{i}\big( q_{j-1}r_{j-1} - q_{j}r_{j}\big)
\end{equation}
for odd $j$.
Therefore the first current reads
\begin{equation} \label{A.19} 
J_j^{[1]} =   \mathrm{i}\big( - \rho_{j-1}^2 q_{j-2}r_{j} +q_{j-1}r_{j-1}\big).
\end{equation}
This computation illustrates the difficulties when taking the $P\to 0$ limit. To reach Eq. \eqref{A.19} still requires explicit cancellations so to yield the actual current.
For general $n$, one has to  rely on an abstract argument.\bigskip\\
\textit{Existence of local currents}. We want to establish that  there is a local current local, $J^{[n]}_{j}$, such that
\begin{equation} \label{A.20}
\{(C^n)_{j,j},H\} =J^{[n]}_{j} -  J^{[n]}_{j+1}.
\end{equation}

We fix $n$ and choose $N > 4n$. $\{Q^{[n]}_0,H\}$ is a polynomial of degree at most $2n+2$. This polynomial is decomposed into \textit{patterns} consisting of monomials and their spatial shifts, denoted by $\omega_j$. An example would be $\omega_j = r_{j-1}(q_{j+3})^2$. Then $\{Q^{[n]}_0,H\}$ is a sum of terms of the form
\begin{equation} \label{A.21}
\sum_{|\ell| \leq 2n}a^{(\ell)} \omega_{\ell}
\end{equation}
with some complex coefficients $a^{(\ell)}$, which may vanish. Since $\{Q^{[n]}_j,H\}$ is shift covariant, from the conservation law,
\begin{equation} \label{A.22}
\sum_{j=0}^{N-1}\sum_{|\ell| \leq 2n+2}a^{(\ell)} \omega_{\ell +j} = 0.
\end{equation}
Relabelling the sum over $j$, one arrives at
\begin{equation} \label{A.23}
\sum_{|\ell| \leq 2n+2}a^{(\ell)} = 0.
\end{equation}
To have a one-shift covariant current density means
 \begin{equation} \label{A.24}
 \sum_{|\ell| \leq 2n+2} b^{(\ell)}(\omega_\ell - \omega_{\ell+1})= J^\omega_0 - J^\omega_{1}.
 \end{equation}
Using \eqref{A.23}, the coefficients $b^{(\ell)}$ are uniquely determined through the $a^{(\ell)}$'s. The total fields $Q^{[n],N}$ are unique,
 but the one-shift covariant local densities  $Q^{[n],N}_j$ are not. Once the densities are fixed,  
 the corresponding one-shift covariant current density is determined.\bigskip\\
\textbf{Acknowledgements}. Highly appreciated are illuminating discussions with Tamara Grava, Guido Mazzuca, and Gaultier Lambert.
This manuscript was completed during the program on ``Universality and Integrability in Random Matrix Theory and Interacting Particle Systems" at the MSRI, Berkeley. I am most grateful for the splendid hospitality.

\end{document}